\documentclass[aps,pra,prabib,twocolumn,showpacs]{revtex4}
\usepackage{amsmath}
\begin{document}
\title{The unitary gas in an isotropic harmonic trap:
symmetry properties and applications}

\author{F\'{e}lix Werner}
\affiliation{Laboratoire Kastler Brossel, \'Ecole Normale
Sup\'erieure, 24 rue Lhomond, 75231 Paris Cedex 05, France}
\author{Yvan Castin}
\affiliation{Laboratoire Kastler Brossel, \'Ecole Normale
Sup\'erieure, 24 rue Lhomond, 75231 Paris Cedex 05, France}

\date{\today}
\begin{abstract}
We consider $N$ atoms trapped in an isotropic harmonic potential,
with $s$-wave interactions of infinite scattering length.
In the zero-range limit, we obtain several exact analytical results:
mapping between the trapped problem and the free-space zero-energy
problem,
separability in hyperspherical coordinates, $SO(2,1)$ hidden symmetry,
and relations between the moments of the trapping potential energy and
the moments of the total energy.
\end{abstract}

\maketitle

Strongly interacting degenerate Fermi gases with two spin components
are studied in present experiments with ultra-cold atoms
\cite{experiments}: by tuning the interaction
strength between the atoms of different spin states {\it via} a
Feshbach resonance, one can even reach the so-called unitary limit
\cite{theory} where the interaction strength in the $s$-wave channel
reaches the maximal amplitude
allowed by quantum mechanics in a gas. More precisely, this means that
the $s$-wave scattering
amplitude between two particles reaches the value
\begin{equation}
f_k = -\frac{1}{ik}
\label{eq:unitarity}
\end{equation}
for the relative momenta $k$ that are relevant in the gas, in
particular for $k$ of the order
of the Fermi momentum $k_F$ of the particles. This implies that the
$s$-wave scattering length $a$
is set to infinity (which is done in practice by tuning an external
magnetic field).
This also implies that $k |r_e|\ll 1$, where $r_e$ is the effective
range of the interaction potential,
a condition well satisfied in present experiments on broad Feshbach
resonances.

The maximally-interacting gas defined by these conditions is called
the unitary quantum gas \cite{theory}. 
It has universal properties since all the details
of the interaction have dropped out of the problem.
Theoretically, for spin 1/2 fermions with equal populations in the two
spin states,
equilibrium properties have been calculated in the thermodynamical
limit in the spatially
homogeneous case using Monte-Carlo methods; at finite temperature
\cite{Svistunov,Bulgac}, and at zero temperature with a fixed node
approximation \cite{fixed_node}.
In practice, the unitary gases produced experimentally are stored in
essentially harmonic traps, which rises
the question of the effect of such an external potential. In this
paper, we consider a specific
aspect of this question: restricting to perfectly isotropic harmonic
traps, but with no
constraint on the relative spin populations, we show
that the unitary quantum gas admits interesting symmetry properties
that have measurable
consequences on its spectrum and on the many-body wavefunctions. These
properties imply that 
there is a mapping between the $N$-body eigenfunctions in a trap and
the zero-energy $N$-body eigenfunctions in free space;
the $N$-body problem is separable in hyperspherical coordinates; and
there exist relations
between the moments of the trapping potential energy and those of the total
energy at thermal
equilibrium.

A unitary {\it Bose} gas was not produced yet. This is related to the
Efimov effect \cite{Efimov}:
when three bosons interact with a short range potential of infinite
scattering length, an effective three-body
attraction takes place, leading in free space to the existence of
weakly bound trimers.
This effective attraction generates high values of $k$ so that the 
unitarity condition Eq.(\ref{eq:unitarity}) is violated.
It also gives a short lifetime to the gas by activating
three-body losses due to the formation of deeply bound molecules
\cite{manips_bosons,grimm_efimov,pertes_braaten}.
In an isotropic harmonic trap, for three bosons, there exist efimovian
states
\cite{Pethick,Koehler}, but there also exist eigenstates not
experiencing the Efimov effect \cite{Pethick,Werner}. These last
states are universal
(in the sense that they depend only on $\hbar$, the mass $m$ and the
trapping frequency $\omega$) and
they are predicted to be long-lived \cite{Werner}. 
The results of the present paper apply to all universal states,
fermionic or bosonic, but do not apply to the efimovian states.

\section{Our model for the unitary gas}

The physical system considered in this paper is a set of $N$ particles
of equal mass $m$
(an extension to different
masses is given in Appendix \ref{appen:masses}). The particles are of
arbitrary spin and follow arbitrary statistics;
the Hamiltonian is supposed to be spin-independent so that the
$N$-body wavefunction
$\psi$ that we shall consider corresponds to a given spin
configuration \cite{aide}.
The particles are trapped by the same isotropic harmonic potential of
frequency $\omega$.
We collect all the positions $\vec{r}_i$ of the particles in a single
$3N$ component vector:
\begin{equation}
\vec{X}\equiv(\vec{r}_1,\ldots,\vec{r}_N).
\label{eq:X}
\end{equation}
Its norm
\begin{equation}
X=\|\vec{X}\|=\sqrt{\sum_{i=1}^N r_i^2}
\label{def_X}
\end{equation}
is called the hyperradius.
We will also use the unit vector
\begin{equation}
\vec{n} \equiv \vec{X} / X
\label{def_n}
\end{equation}
(which may be parametrized by $3N-1$ hyperangles).
The coordinates $(X,\vec{n})$ are called hyperspherical coordinates
\cite{Esry}.
The total trapping potential energy simply writes
\begin{equation}
H_{\rm trap}=\frac{1}{2} m \omega^2 X^2.
\label{eq:def_H_trap}
\end{equation}

The interaction between the particles is assumed to be at the
unitary limit defined 
in Eq.(\ref{eq:unitarity}); one can then replace the interaction by
contact
conditions on the $N$-body wavefunction (see
e. g. \cite{Olshanii,Shlyap}
and references therein):
when the distance $r_{ij}=\|\vec{r}_j-\vec{r}_i\|$ between particles
$i$ and $j$ tends to zero,
there exists a function $A$ such that
\begin{equation}
\psi(\vec{X}\,) \underset{r_{ij}\rightarrow 0}{=}
\frac{A(\vec{R}_{ij},\{\vec{r}_k:k\neq i,j\})}{r_{ij}}+O(r_{ij})
\label{eq:cc}
\end{equation}
where $\vec{R}_{ij}=(\vec{r}_i+\vec{r}_j)/2$ is the fixed
center-of-mass position of particles $i$ and $j$,
and $\{\vec{r}_k:k\neq i,j\}$ are the positions of the other
particles.
When none of the particle positions coincide, the stationary
wavefunction $\psi$
solves Schr\"odinger's equation, $H \psi = E \psi$,
with the Hamiltonian
\begin{equation}
H = -\frac{\hbar^2}{2m} \Delta_{\vec{X}} + \frac{1}{2} m\omega^2 X^2.
\end{equation}
This model is expected to be exact for universal states in the limit
of 
a zero range of the interaction potential \cite{Olshanii}.

\section{Scaling properties of the trapped unitary gas}

\subsection{What is scale invariance ?}

A fundamental property of the contact conditions Eq.(\ref{eq:cc}) is
their invariance
by a rescaling of the spatial coordinates. More precisely, we define a
rescaled wavefunction
$\psi_\lambda$ by
\begin{equation}
\psi_\lambda(\vec{X}\,) \equiv \psi(\vec{X}/\lambda)
\end{equation} 
where $\lambda>0$ is the scaling factor. Then, if $\psi$ obeys the
contact conditions, so does $\psi_\lambda$
for any $\lambda$.
Note that this property holds only because the scattering length is
infinite (for a finite value
of $a$, $1/r_{ij}$ in Eq.(\ref{eq:cc}) would be replaced by
$1/r_{ij}-1/a$, which breaks scale invariance).

In free space (that is
for $\omega=0$), the scale
invariance implies the following property:
if $\psi$ is an eigenstate of energy $E$,
then $\psi_\lambda$
is an eigenstate of energy $E/\lambda^2$ for any $\lambda$
\cite{felix}.
This implies the absence of bound states in
free space: otherwise the scaling transform would generate a continuum
of states which are square integrable
(after elimination of the center of mass variables), and this is
forbidden for
a Hermitian problem \cite{HF}.

When $E=0$, one finds (see Appendix \ref{appen:scale}) that the free
space
eigenstates can be assumed to be scale-invariant, i.e. 
there exists an exponent $\nu$ such that
\begin{equation}
\psi_\lambda(\vec{X}\,) = \lambda^{-\nu} \psi(\vec{X}).
\label{eq:si}
\end{equation}
Taking the derivative of this relation with respect to $\lambda$ in
$\lambda=1$, this shows that
$\psi$ is an eigenstate of the dilatation operator,
\begin{equation}
\hat{D} \equiv \vec{X}\cdot\partial_{\vec{X}},
\end{equation}
with the eigenvalue $\nu$. This result is interesting for section
\ref{sec:mapping}.

The presence of a harmonic trap introduces the harmonic oscillator
length scale
$a_{\rm ho}=\sqrt{\hbar/m\omega}$, so that the eigenstates cannot be
scale-invariant as in Eq.(\ref{eq:si}).
However, if $\psi$ obeys the contact condition, so do the
$\psi_\lambda$'s: as we shall see, this allows
to identify general properties of the eigenstates in the trap.

\subsection{Scaling solution in a time dependent trap}

We now assume that the trap frequency $\omega$, while keeping a fixed
value $\omega(0)$
for all times $t<0$, has an arbitrary time dependence at positive
times.
Let us assume that, at $t\le0$, the system is in a stationary state of
energy $E$. Then at positive times
the wavefunction of the system will be deduced from the $t=0$
wavefunction  by the combination
of gauge and scaling transform \cite{CRAS}:
\begin{equation}
\psi(\vec{X},t) = \frac{e^{-i E \tau(t)/\hbar}}{\lambda(t)^{3N/2}}
e^{i m X^2 \dot{\lambda}(t)/2\hbar \lambda(t)}
\psi(\vec{X}/\lambda(t),0)
\label{eq:ss}
\end{equation}
where the time dependent scaling parameter obeys the Newton-like
equation
\begin{equation}
\ddot\lambda = \frac{\omega^2(0)}{\lambda^3} -\omega^2(t) \lambda
\label{eq:newton}
\end{equation}
with the initial conditions $\lambda(0)=1, \dot\lambda(0) = 0$.
We also introduced an effective time $\tau$ given by
\begin{equation}
\tau(t) = \int_0^t \frac{dt'}{\lambda^2(t')}.
\end{equation}

This result may be extended to an arbitrary initial state as follows:
\begin{equation}
\psi(\vec{X},t) = \frac{1}{\lambda(t)^{3N/2}}
e^{i m X^2 \dot{\lambda}(t)/2\hbar \lambda(t)}
\tilde{\psi}(\vec{X}/\lambda(t),\tau(t)).
\end{equation}
where $\tilde{\psi}$ evolves with the $t<0$ Hamiltonian (i.e. in the
unperturbed trap of frequency $\omega(0)$).

As shown by Rosch and Pitaevskii \cite{Rosch}, the existence of such a
scaling and gauge time dependent
solution is related to a SO(2,1) hidden symmetry of the problem. This
we rederive in the next subsection.

\subsection{Raising and lowering operators, and SO(2,1) hidden
  symmetry}

We consider the following gedanken experiment: one perturbs the gas
in an infinitesimal way by modifying the trap frequency in a time
interval $0<t<t_f$. After the excitation period ($t>t_f$), the trap
frequency assumes its initial value
$\omega(0)$. The scaling parameter then slightly deviates from unity,
$\lambda(t)=1+\delta\lambda(t)$ with $|\delta\lambda|\ll
1$. Linearizing the equation of motion Eq.(\ref{eq:newton})
in $\delta\lambda$, one finds that $\delta\lambda$ oscillates as
\begin{equation}
\delta\lambda(t) = \epsilon e^{-2i\omega t} + \epsilon^* e^{2i\omega
  t},
\label{eq:dl}
\end{equation}
where we set $\omega=\omega(0)$ to simplify the notation.
The gedanken experiment has therefore excited an undamped breathing
mode of frequency $2\omega$ \cite{Rosch}.

We now interpret this undamped oscillation in terms of a property of
the $N$-body spectrum of the system.
Expanding Eq.(\ref{eq:ss}) to first order in $\delta\lambda(t)$ leads
to
\begin{eqnarray}
\psi(\vec{X},t) &=& e^{i\alpha}\left[e^{-i E t/\hbar} -
\epsilon e^{-i(E+2\hbar\omega)t/\hbar} L_+
\right.
\nonumber
\\ & + & \left.
\epsilon^* e^{-i(E-2\hbar\omega)t/\hbar} L_-\right]
\psi(\vec{X},0)+O(\epsilon^2)
\end{eqnarray}
(the phase $\alpha$ depends on the details of the excitation
procedure).
This reveals that the initial stationary state $E$ was coupled by the
excitation procedure 
to other stationary states of energies $E\pm 2\hbar\omega$.
Remarkably, the wavefunction of these other states can be obtained
from the initial one
by the action of raising and lowering operators:
\begin{eqnarray}
L_+ &=& +\frac{3N}{2} +\hat{D} + \frac{H}{\hbar\omega} -m\omega
X^2/\hbar
\label{L_+}
\\
L_- &=& -\frac{3N}{2} -\hat{D} + \frac{H}{\hbar\omega} -m\omega
X^2/\hbar.
\end{eqnarray}
Repeated action of $L_+$ and $L_-$ will thus generate a ladder of
eigenstates with regular
energy spacing $2\hbar\omega$.

The hidden SO(2,1) symmetry of the problem then results from the fact
that
$H$, $L_+$ and $L_-$ have commutation relations equal (up to numerical
factors)
to the ones of the Lie algebra of the SO(2,1) group, as was checked in
\cite{Rosch}:
\begin{eqnarray}
[H,L_+] &=&  2\hbar\omega L_+\\
\left[H,L_-\right] &=&  -2\hbar\omega L_-\\
\left[L_+,L_-\right] &=&  -4 \frac{H}{\hbar\omega}
\end{eqnarray}

From the general theory of Lie algebras, one may form the so-called
Casimir operator
which commutes with all the elements of the algebra, that is with $H$
and $L_\pm$; it is
given by \cite{Rosch}:
\begin{equation}
\hat{C} = H^2 - \frac{1}{2}  (\hbar\omega)^2 (L_+ L_- + L_- L_+).
\end{equation}
Consider a ladder of eigenstates; as we will show later, the
hermiticity of $H$
implies that this ladder has a ground energy step, of value
$E_0$. Within this ladder,
the Casimir invariant assumes a constant value, 
\begin{equation}
C= E_0 (E_0-2\hbar \omega).
\end{equation}

\subsection{Virial theorem}

Another application of the existence of raising and lowering operators
is
the virial theorem for the unitary gas.
For a given eigenstate of $H$ of energy $E$ and real wavefunction
$\psi$, $L_-|\psi\rangle$ is either
zero (if $\psi$ is the ground step of a ladder) or an eigenstate of
$H$ with a different
energy. Assuming that $H$ is hermitian, this implies $\langle\psi|
L_-|\psi\rangle=0$,
and leads to \cite{virial}:
\begin{equation}
\left<\psi|H|\psi\right>=2\left<\psi|H_{\rm trap}|\psi\right>.
\label{eq:virial_psi}
\end{equation}
At thermodynamical equilibrium, one thus has
\begin{equation}
\left<H\right>=2\left<H_{\rm trap}\right>,
\end{equation}
that is the total energy is twice the mean trapping potential energy.

This virial theorem is actually also valid for an anisotropic harmonic
trap (this result is due to Fr\'ed\'eric
Chevy). One uses the Ritz theorem, stating that an eigenstate of a
hermitian Hamiltonian is a stationary
point of the mean energy. As a consequence, the function of $\lambda$
\begin{eqnarray}
E(\lambda) &\equiv& \frac{\langle\psi_\lambda|H|\psi_\lambda\rangle}
{\langle\psi_\lambda|\psi_\lambda\rangle}
\nonumber
\\
&=&
\lambda^{-2}\left<\psi|H-H_{\rm trap}|\psi\right> +
\lambda^{2}\left<\psi|H_{\rm trap}|\psi\right>
\end{eqnarray}
satisfies $(dE/d\lambda)(\lambda=1)=0$, which leads to the virial
theorem. This relies simply on the scaling
properties of the harmonic potential, irrespective of its isotropy.

The proportionality  between $\left<H\right>$ and $\left<H_{\rm
    trap}\right>$
resulting from the virial theorem was checked experimentally
    \cite{thomas_virial}.

\section{Mapping to zero-energy free-space eigenstates}
\label{sec:mapping}

Usually, the presence of a harmonic trap in the experiment makes the
theoretical analysis more
difficult than in homogenenous systems. Here we show that, remarkably,
the case of an isotropic
trap for the unitary gas can be mapped exactly to the zero-energy free-space problem (which remains,
of course, an unsolved many-body problem).

More precisely, all the universal $N$-body eigenstates can be put in
the unnormalized form:
\begin{equation}
|\psi_{\nu,q}\rangle =  \left(L_+\right)^q \, e^{-\hat{X}^2/2 a_{\rm
 ho}^2} 
|\psi^0_\nu\rangle
\label{eq:mapping}
\end{equation}
and have an energy
\begin{equation}
E_{\nu,q} = (\nu +2q +3N/2) \hbar\omega
\label{eq:E_vs_nu}
\end{equation}
where $q$ is a non-negative integer, $L_+$ is the raising operator
defined
in Eq.(\ref{L_+}),
and $\psi^0_\nu$ is a zero-energy eigenstate of the free-space problem
which is scale-invariant:
\begin{equation}
\psi^0_\nu(\vec{X}/\lambda)= 
\psi^0_\nu(\vec{X}\,)/ \lambda^{\nu} 
\end{equation}
for all real scaling parameter $\lambda$, $\nu$ being the real scaling
exponent \cite{abus}.

We also show that the reciprocal is true, that is each zero-energy
free-space
eigenstate which is scale-invariant with a real exponent $\nu$
generates
a semi-infinite ladder of eigenstates in the trap, according to
Eq.(\ref{eq:mapping},\ref{eq:E_vs_nu}).

We note that Eq.(\ref{eq:E_vs_nu}) generalizes to excited states
a relation obtained in \cite{Tan} for the many-body ground state.

\subsection{From a trap eigenstate to a free-space eigenstate}

We start with an arbitrary eigenstate in the trap.
By repeated action of $L_-$ on this eigenstate, we produce a sequence
of eigenstates of decreasing energies.
According to the virial theorem Eq.(\ref{eq:virial_psi}), the total energy of a universal state
is positive, since
the trapping potential energy is positive. This means that the
sequence produced above terminates.
We call $\psi$ the last non-zero wavefunction of the sequence, an
eigenstate of $H$ with energy
$E$ that satisfies
$L_- |\psi\rangle=0$.  To integrate this equation, we use the
hyperspherical coordinates $(X,\vec{n})$
defined in Eq.(\ref{def_X},\ref{def_n}).
Noting
that the dilatation operator
is simply $\hat{D}=X\cdot\partial_X$ in hyperspherical coordinates, we
obtain:
\begin{equation}
\psi(\vec{X}\,) = e^{-X^2/2a_{\rm ho}^2} \, X^{E/(\hbar\omega)-3N/2}
f(\vec{n}).
\label{eq:sol_L-}
\end{equation}
Then one defines 
\begin{equation}
\psi^0(\vec{X}) \equiv e^{X^2/2a_{\rm ho}^2}  \psi(\vec{X}\,).
\label{eq:def_psi0}
\end{equation}
One checks that this wavefunction obeys the contact conditions
 Eq.(\ref{eq:cc}), since
$X^2$ varies quadratically with $r_{ij}$ at fixed $\vec{R_{ij}}$ and
 $\{\vec{r}_k,k\neq i,j\}$.
$\psi^0$ is then found to be a zero-energy eigenstate in free space,
 by direct insertion into Schr\"odinger's equation. 
But one has also from Eq.(\ref{eq:sol_L-},\ref{eq:def_psi0}):
\begin{equation}
\psi^0(\vec{X})=X^{E/(\hbar\omega)-3N/2} f(\vec{n}),
\end{equation}
so that $\psi^0$ is scale-invariant, with a real exponent $\nu$
related to the energy $E$ by Eq.(\ref{eq:E_vs_nu}).
This demonstrates Eq.(\ref{eq:mapping},\ref{eq:E_vs_nu}) for $q=0$,
that is for the ground step of each ladder.

One just has to apply a repeated action of the raising operator $L_+$
on the ground step wavefunction
to generate a semi-infinite ladder of eigenstates: this corresponds to
$q>0$ in Eq.(\ref{eq:mapping},\ref{eq:E_vs_nu}).
Note that the repeated action of $L_+$ cannot terminate since
$L_+|\psi\rangle = 0$ for a non-zero $\psi$ 
implies that $\psi$ is not square-integrable.

\subsection{From a free-space eigenstate to a trap eigenstate}

The reciprocal of the previous subsection is also true:
starting from an arbitrary zero-energy free-space eigenstate that is
scale-invariant,
one multiplies it by the Gaussian factor $\exp(-X^2/2a_{\rm ho}^2)$,
and one checks that the resulting wavefunction is an eigenstate of the
Hamiltonian of the trapped system, 
obeying the contact conditions \cite{nu_efi}.
Applying $L_+$ then generates the other trap eigenstates.

\subsection{Separability in hyperspherical coordinates}
\label{subsec:joli}

Let us reformulate the previous mapping using the hyperspherical
coordinates
$(X,\vec{n})$
defined in Eq.(\ref{def_X},\ref{def_n}).
A free-space scale-invariant zero-energy eigenstate takes the form
$\psi^0(\vec{X}\,) = X^\nu f_\nu(\vec{n}\,)$, and the universal
eigenstates in the trap
have an unnormalized wavefunction
\begin{equation}
\psi_{\nu,q}(\vec{X}\,) = X^\nu e^{-X^2/2a_{\rm ho}^2}
L_q^{(\nu-1+3N/2)}(X^2/a_{\rm ho}^2)\,
f_\nu(\vec{n}\,),
\label{eq:joli}
\end{equation}
where $L_q^{(.)}$ is the generalized Laguerre polynomial of degree
$q$. This is obtained from
the repeated action of $L_+$ in Eq.(\ref{eq:mapping}) and from the
recurrence relation
obeyed by the Laguerre polynomials:
\begin{equation}
(q+1) L_{q+1}^{(s)}(u) -(2q+s+1-u) L_q^{(s)}(u) + (q+s)
L_{q-1}^{(s)}(u)=0.
\label{eq:recur} 
\end{equation}
We have thus separated out the hyperradius $X$ and the hyperangles
$\vec{n}$. The hyperangular
wavefunctions $f_\nu(\vec{n}\,)$  and the exponents $\nu$ are not
known for $N\geq 4$.
However, we have obtained the hyperradial wavefunctions, i.e.\ the $X$
dependent part of the many-body wavefunction.
A more refined version of these separability results can be obtained
by first separating out the center of mass (see Appendix
\ref{appen:internal}),
but this is not useful for the next Section.

\section{Moments of the trapping potential energy}
\subsection{Exact relations}
As an application of the above results,  we now obtain the following
exact relations on the statistical properties
of the trapping potential energy, relating its moments to the moments
of the full energy,
when the gas is at thermal equilibrium.
For the definition of the trapping potential energy, see
Eq.(\ref{eq:def_H_trap}).

At zero temperature, its moments as a function of the ground
state energy $E_0$ are given by:
\begin{equation}
\langle \left(H_{\rm trap}\right)^n \rangle=
E_0\big(E_0+\hbar\omega\big) \ldots
  \big(E_0+(n-1)\hbar\omega\big) /2^n.
\label{moment_T=0}
\end{equation}
At finite temperature $T$, 
the first moment is given by the virial theorem
\begin{equation}
\langle H_{\rm trap}\rangle=\langle H\rangle / 2
\label{moment1}
\end{equation}
and the second moment by
\begin{equation}
\langle \left(H_{\rm trap}\right)^2\rangle = \left[\langle H^2\rangle
  +
  \langle
  H\rangle \hbar\omega\cdot
\mbox{cotanh}\left(\frac{\hbar\omega}{k_B T}\right) \right]/4.
\label{moment2}
\end{equation}

\subsection{Derivation}
The zero temperature result Eq.(\ref{moment_T=0}) follows directly
from Eq.(\ref{eq:joli}): for $q=0$, the Laguerre polynomial is
constant so that
the probability distribution of $X$ is a power law times a Gaussian;
the moments are then given by integrals
that can be expressed in terms of the $\Gamma$ function.

For finite $T$, the idea of our derivation is the following:
the hyperradial part of the $N$-body wavefunction $\psi_{\nu,q}$ is
known from
Eq. (\ref{eq:joli});
and thus the probability distribution of
$X$ in the state $\left|\psi_{\nu,q}\right>$
is known, in terms of $\nu,q$.
While the thermal distribution of $q$ is simple, the one of $\nu$ is
not, but
$\nu$ is related to the total energy by Eq.(\ref{eq:E_vs_nu}).

We will need the intermediate quantities:
\begin{equation}
B_{n,p} \equiv \frac{\int_0^\infty du\, e^{-u} u^{s+n}
  L_{q+p}^{(s)}(u) L_q^{(s)}(u)}
{\int_0^\infty du\, e^{-u} u^{s} \left[L_q^{(s)}(u)\right]^2},
\end{equation}
where $s\geq 0$; $n,q$ are non-negative integers; and $p$ is an
integer of arbitrary 
sign. These quantities can be calculated 
with the $n=0$ `initial' condition $B_{0,p}=\delta_{0,p}$ and the
recurrence relation
\begin{eqnarray}
B_{n+1,p} &=& -(q+p+1) B_{n,p+1} +\left[2(q+p)+s+1\right]B_{n,p} 
\nonumber
\\
& &-(q+p+s) B_{n,p-1}
\end{eqnarray}
which follows from the recurrence relation Eq.(\ref{eq:recur}) on
Laguerre polynomials.

This allows to calculate the moments of the trapping energy in the
step
$q$ of a ladder of exposant $\nu$, using Eq.(\ref{eq:joli}):
\begin{equation}
\frac{\langle\psi_{\nu,q}| X^{2n}|
  \psi_{\nu,q}\rangle}{\langle\psi_{\nu,q}|\psi_{\nu,q}\rangle}=
B_{n,0} \, a_{\rm ho}^{2n}.
\end{equation}
Here we have set
\begin{equation}
s=\nu-1+3N/2
\end{equation}
in accordance with Eq.(\ref{eq:joli}).

Assuming thermal equilibrium in the canonical ensemble, the thermal
average can be performed
over the statistically independent variables $q$ and $s$. The moments
of $q$ are easy to calculate,
because of the ladder structure with equidistant steps:
\begin{equation}
\langle q^n \rangle = \frac{\sum_{q=0}^{+\infty} q^n
  e^{-q\hbar\omega/k_B T}}{\sum_{q=0}^{+\infty} e^{-q\hbar\omega/k_B
  T}}.
\end{equation}
The moments of $s$ are not known exactly but they can be eliminated in
terms of the moments of the total energy
$E$ and of the moments of $q$ using the relation
$E=(s+1+2q)\hbar\omega$.
This leads to the exact relations (\ref{moment1},\ref{moment2}).
This method in principle allows to calculate relations for moments of
arbitrary given order, but the algebra
becomes cumbersome.

\section{Conclusion}
In this paper we have derived several exact properties of the unitary
gas in an isotropic harmonic trap.
The spectrum is formed of ladders; the steps of a ladder are spaced by an energy $2\hbar
\omega$, and linked by
raising and lowering operators. This allows to map the trapped problem
to the free-space one.
The problem is separable in hyperspherical coordinates.
This allows to derive exact relations between the moments of the
trapping potential energy and the moments of the total energy.
The relation between the first moments is the virial theorem;
the relation between the second moments may be useful for thermometry,
as will be studied elsewhere.

We thank F. Chevy, J. Dalibard, S. Nascimb\`{e}ne, L. Pricoupenko,
J. Thomas and C.-F. Vergu for
very useful discussions.
LKB is a {\it Unit\'{e} de Recherche de l'ENS et de l'Universit\'{e}
  Paris 6, associ\'{e}e au CNRS}.
Our research group is a member of IFRAF.

\appendix

\section{Extension to particles with different masses}
\label{appen:masses}
All our results remain valid if the particles have different masses
$m_1,\ldots,m_N$; provided that the trapping frequency $
\omega$ remains the
same for all the particles.
We define a mean mass:
\begin{equation}
m\equiv\frac{m_1+\ldots+m_N}{N}.
\end{equation}
The definition of $\vec{X}$ and $X$, given by
Eq.(\ref{eq:X},\ref{def_X}) for equal
masses, has to
be generalized to:
\begin{eqnarray}
\vec{X}&\equiv&\left(\sqrt{\frac{m_1}{m}}\vec{r}_1,\ldots,\sqrt{\frac{m_N}{m}}\vec{r}_N\right).
\\
X&\equiv&\|\vec{X}\|=\sqrt{\sum_{i=1}^N \frac{m_i}{m} r_i^2}.
\end{eqnarray}
With this new definition of $X$, the trapping potential energy is
still given by Eq.(\ref{eq:def_H_trap}).

In the definition of the zero-range model, the contact conditions
Eq.(\ref{eq:cc}) remain unchanged, except that the fixed center of
mass position of particles
$i$ and $j$ is now
$\vec{R}_{ij}\equiv(m_i\vec{r_i}+m_j\vec{r_j})/(m_i+m_j)$.

In Appendix \ref{appen:internal}, the center of mass position has to
be redefined as
\begin{equation}
\vec{C}=\frac{\left(m_1 \vec{r_1}+\ldots+m_N
\vec{r_N}\right)}{(m_1+\ldots+m_N)}
\end{equation}
and the internal hyperangular coordinates become:
\begin{eqnarray}
R&=&\sqrt{\sum_{i=1}^{N} \frac{m_i}{m}(\vec{r}_i-\vec{C}\,)^2}
\\
\vec{\Omega}&=&\left(\sqrt{\frac{m_1}{m}}\frac{\vec{r}_1-\vec{C}}{R},\ldots,\sqrt{\frac{m_N}{m}}\frac{\vec{r}_N-\vec{C}}{R}\right).
\end{eqnarray}

With these modified definitions, all the results of this paper remain
valid.

\section{Scale invariance of the zero-energy free-space eigenstates}
\label{appen:scale}

In this appendix, we show that the zero-energy free-space eigenstates
of the
Hamiltonian may be chosen as being scale-invariant, that is as
eigenstates of
the dilatation operator $\hat{D}$, under conditions ensuring the
hermiticity
of the Hamiltonian.

Consider the zero-energy eigensubspace of the free-space Hamiltonian.
This subspace is stable under the action of $\hat{D}$.
If one assumes that $\hat{D}$ is diagonalizable within this subspace,
the corresponding eigenvectors form a complete family of scale
invariant zero-energy
states.
If $\hat{D}$ is not diagonalizable, we introduce the Jordan normal
form of $\hat{D}$.

Let us start with the case of a Jordan normal form of dimension two,
written
as
\begin{equation}
\mbox{Mat}(\hat{D})= \left(
\begin{array}{cc}
\nu & 1 \\
0   & \nu 
\end{array}
\right),
\end{equation}
in the sub-basis $|e_1\rangle, |e_2\rangle$.
The ket $|e_1\rangle$ is an eigenstate of $\hat{D}$ with the
eigenvalue $\nu$.
We assume that the center of mass motion is at rest, with no loss of
generality since it is separable in free space.
Using the internal hyperspherical coordinates $(R,\vec{\Omega}\,)$
defined in Appendix
\ref{appen:internal}, we find that $\hat{D}$ reduces to
the operator $R\partial_R$.  Integrating $R\partial_R e_1 = \nu e_1$
leads to
\begin{equation}
e_1(\vec{X})= R^\nu \phi_1(\vec{\Omega}\,).
\end{equation}
The ket $|e_2\rangle$ is not an eigenstate of $\hat{D}$ but obeys
$R\partial_R e_2 = \nu e_2 + e_1$, which, after integration, gives
\begin{equation}
e_2(\vec{X}\,) = R^\nu \log R \phi_1(\vec{\Omega}\,) + R^\nu
\phi_2(\vec{\Omega}).
\label{eq:e2}
\end{equation}
One can assume that $\phi_1$ and $\phi_2$ are orthogonal on the unit
sphere (by redifining $e_2$ and $\phi_2$). 
It remains to use the fact that both $e_1$ and $e_2$ are zero-energy
free-space eigenstates.
From the form of the Laplacian in hyperspherical
coordinates in $d=3N-3$ dimensions, see Eq.(\ref{eq:Hint}), the
condition $\Delta_{\vec{X}} e_1=0$ leads to 
\begin{equation}
T_{\vec{\Omega}} \phi_1 = - \nu(\nu+d-2)\phi_1.
\end{equation}
The condition $\Delta_{\vec{X}} e_2=0$ then gives $T_{\vec{\Omega}}
\phi_2=-\nu(\nu+d-2)\phi_2-(2\nu+d-2)\phi_1$, which leads to the
constraint \cite{why_s=0}:
\begin{equation}
\nu = 1-d/2.
\label{eq:s=0}
\end{equation}
At this stage, for this 'magic' value of $\nu$, it seems that there
may exist
non scale-invariant zero-energy eigenstates.

To proceed further, one has to check for the hermiticity of the free
space 
Hamiltonian. This requires a reasoning at arbitrary, non zero energy.
We use the fact that the following wavefunction obeys the contact
conditions,
\begin{equation}
\psi(\vec{X}\,) = u(R) R^\nu \phi_1(\vec{n}\,),
\end{equation}
where $u(R)$ is a fonction with no singularity, except maybe in $R=0$
\cite{lemma}.
Using again the expression of the Laplacian in internal hyperspherical
coordinates,
one finds that $\psi$ is an eigenstate of the free-space Hamiltonian
if
$u(R)$ is an eigenstate of
\begin{equation}
\hat{h} = -\frac{\hbar^2}{2m} (\partial_R^2 + R^{-1}\partial_R).
\end{equation}
One checks that Hermiticity of the free-space Hamiltonian for the
wavefunction
$\psi$ implies hermiticity of $\hat{h}$ for the `wavefunction' $u(R)$.
Note that $\hat{h}$ is simply the free-space Hamiltonian for 2D
isotropic wavefunctions.
It is hermitian over the domain of wavefunctions $u(R)$ with a
non-infinite
limit in $R=0$. Including the ket $|e_2\rangle$ in the domain of the
$N$-body
free-space Hamiltonian amounts to allowing for `wavefunctions' $u(R)$
that diverge
as $\log\, R$ for $R\rightarrow 0$: this breaks the Hermiticity of
$\hat{h}$, since
this leads to a (negative energy) continuum of square integrable
eigenstates of
$\hat{h}$,
\begin{equation}
u_\kappa(R) = K_0(\kappa R)
\end{equation}
with eigenenergy $-\hbar^2 \kappa^2/2m$, for all $\kappa>0$. Here
$K_0(x)$
is a modified Bessel function of the second kind.
Hermiticity may be restored by a filtering of this contiunuum
\cite{Morse},
adding the extra contact condition $u(R)=\log(R/l) + o(1)$ for
$R\rightarrow 0$,
but the introduction of the fixed length $l$ breaks the universality
of the
problem and is beyond the scope of this paper
We thus exclude $e_2$ from
the domain of the Hamiltonian.

This discussion may be extended to Jordan forms of higher order.
For example, a Jordan form of dimension $3$ generates a ket
$|e_3\rangle$
such that $(\hat{D}-\nu) e_3=e_2$. But $e_2$ must be excluded from the
domain of the Hamiltonian by the above reasoning.
Since we want the domain to be stable under $\hat{D}$, $e_3$ must be
excluded as well.

As a conclusion, to have a free-space $N$-body Hamiltonian
that is both hermitian and universal (i.e.\ with a scale-invariant
domain) 
forces to reject the non scale-invariant zero-energy eigenstates, of
the form Eq.(\ref{eq:e2}).

\section{Separability in internal hyperspherical coordinates}
\label{appen:internal}

We develop here a refined version of the separability introduced in
subsection
\ref{subsec:joli}. First, we separate out the center of mass
coordinates.
Then we obtain the separability in hyperspherical coordinates relative
to the internal
variables of the gas, which allows to derive an effective repulsive
$N-1$ force 
and to get a lower bound on the energy slightly better than the one
$E> 0$ ensuing from the virial theorem.

Let us introduce the following set of coordinates:
\begin{equation}
\vec{C}=\sum_{i=1}^{N} \vec{r}_i/N
\end{equation}
is the position of the center of mass (CM);
\begin{equation}
R=\sqrt{\sum_{i=1}^N (\vec{r}_i-\vec{C}\,)^2}
\end{equation}
is the internal hyperradius; and
\begin{equation}
\vec{\Omega}=\left(\frac{\vec{r}_1-\vec{C}}{R},\ldots,\frac{\vec{r}_N-\vec{C}}{R}\right)
\end{equation}
is a set of dimensionless internal coordinates that can be
parametrized by $3N-4$ internal hyperangles.
In these coordinates, the Hamiltonian decouples as $H=H_{CM} +
H_{\rm int}$ with
\begin{eqnarray}
H_{CM} &=& -\frac{\hbar^2}{2Nm} \Delta_{\vec{C}} + \frac{1}{2} N m
\omega^2 C^2 \\
H_{\rm int} &=& -\frac{\hbar^2}{2m} \left[\partial_R^2 +
  \frac{3N-4}{R} \partial_R 
+\frac{1}{R^2} T_{\vec{\Omega}} \right]
\nonumber
\\
& & + \frac{1}{2} m\omega^2 R^2
\label{eq:Hint}
\end{eqnarray}
where $T_{\vec{\Omega}}$ is the Laplacian on the unit sphere of
dimension $3N-4$.
The contact conditions do not break the separability of the center of
mass valid in a harmonic
trap, so that the stationary state wavefunction may be taken of the
form
\begin{equation}
\psi(\vec{X}\,) = \psi_{CM}(\vec{C}) \psi_{\rm int}(R,\vec{\Omega}\,).
\end{equation}

One can show \cite{mapping_interne} that 
there is separability in internal hyperspherical coordinates:
\begin{equation}
\psi_{\rm int}(R,\vec{\Omega}\,) = \Phi(R) \phi(\vec{\Omega}\,).
\end{equation}
This form may be injected into the internal Schr\"odinger equation
\begin{equation}
H_{\rm int} \psi_{\rm int} = E_{\rm int} \psi_{\rm int}.
\label{eq:int_pb}
\end{equation}
One finds that $\phi(\vec{\Omega}\,)$ is an eigenstate of
$T_{\vec{\Omega}}$ with an eigenvalue that we call
$-\Lambda$. Note that the contact conditions Eq.(\ref{eq:cc}) put a
constraint on $\phi(\vec{\Omega}\,)$ only \cite{lemma}.
The equation for $\Phi(R)$ reads:
\begin{eqnarray}
-\frac{\hbar^2}{2m}\left(\partial_R^2 +\frac{3N-4}{R}\partial_R\right)
 \Phi
& + & \left(\frac{\hbar^2\Lambda}{2mR^2}+\frac{1}{2}m\omega^2
R^2\right)\Phi
\nonumber
\\
& & = E_{\rm int} \Phi.
\end{eqnarray}

A useful transformation of this equation is obtained by the change of
variable:
\begin{equation}
\Phi(R) \equiv R^{(5-3N)/2} F(R),
\end{equation}
resulting in
\begin{eqnarray}
-\frac{\hbar^2}{2m}\left(\partial_R^2 + \frac{1}{R}\partial_R\right)
 F
&+&\left(\frac{\hbar^2 s_R^2}{2mR^2}+\frac{1}{2}m\omega^2 R^2\right)F 
\nonumber
\\
& & =
E_{\rm int} F(R),
\label{eq:F}
\end{eqnarray}
where $s_R$ is such that
\begin{equation}
s_R^2 = \Lambda + \left(\frac{3N-5}{2}\right)^2.
\end{equation}
Formally, the equation for $F$ is Schr\"odinger's equation for a
particle of zero angular momentum 
moving in 2D in a harmonic potential plus a potential $\propto
s_R^2/R^2$.

For $s_R^2\geq 0$, one can choose $s_R\geq 0$.
Assuming that there is no $N$-body resonance, $F(R)$ is bounded for
$R\rightarrow 0$ \cite{Nbody_res}.
The eigenfunctions of Eq.(\ref{eq:F}) can then be expressed in terms
of the generalized Laguerre polynomials:
\begin{equation}
F(R) = R^{s_R}L_q^{s_R}[R^2/a_{\rm ho}^2]e^{-R^2/2a_{\rm ho}^2}
\label{F_R}
\end{equation}
with the spectrum:
\begin{equation}
E_{\rm int}= (s_R+1+2q) \hbar\omega.
\label{eq:E_int}
\end{equation}
This gives a lower bound on the energy of any universal $N$-body
eigenstate:
\begin{equation}
E \geq \frac{5}{2}\hbar \omega
\label{lower_bound}
\end{equation}
for $N>2$ and in the absence of a $N$-body resonance.

For a complex $s_R^2$, the effective 2D Hamiltonian is not hermitian
and this case
has to be discarded.
For $s_R^2<0$, Whittaker functions are square integrable solutions of
the effective 2D problem
for all values $E_{\rm int}$ so that, again, the problem is not
hermitian. One may add extra boundary conditions
to filter out an orthonormal discrete subset (as was done for
$N=3$ bosons \cite{Danilov,Efimov,Pethick,Werner}) but this breaks the
scaling invariance of the domain
and generates non-universal states beyond the scope of the present
paper.

To make the link with the approach of Section \ref{sec:mapping}, we
note that
\begin{equation}
F(R)=R^{s_R}
\label{F_R_free_space}
\end{equation}
is a solution of the effective 2D problem
(\ref{eq:F}) for $\omega=0,E_{\rm int}=0$. Thus a solution of the internal
problem Eq.(\ref{eq:int_pb}) at zero energy in free
space is given by
\begin{equation}
\psi_{\rm int}(R,\Omega)=R^{(5-3N)/2+s_R}\phi(\vec{\Omega}).
\label{psi_int}
\end{equation}
Multiplying this expression by
$C^l Y_l^m(\vec{C}/C)$, one recovers the $\psi_\nu^0$'s of Section
\ref{sec:mapping}, with
\begin{equation}
\nu=\frac{5-3N}{2}+s_R+l.
\label{nu_s}
\end{equation}


\begin{thebibliography}{99}
\bibitem{aide}
For $N$ spin-1/2 fermions, the spin configuration can be chosen as
$|\uparrow\rangle^{\otimes
  N_{\uparrow}}\otimes|\downarrow\rangle^{\otimes N_\downarrow}$
where the first $N_\uparrow$ factors of the tensorial product are the
spin up state and the $N_\downarrow=
N-N_\uparrow$ remaining factors are the spin down state. In this case,
the wavefunction
$\psi(\mathbf{r}_1,\ldots,\mathbf{r}_N)$ is antisymmetric for the
permutation of the positions of the $N_\uparrow$ first
particles and also antisymmetric for the permutation of the last
$N_\downarrow$ particles.
\bibitem{HF}
This can be seen using the Hellmann-Feynman theorem, valid for
Hermitian Hamiltonians: 
taking the derivative with respect to $\lambda$ of  $\langle
\psi_\lambda
|H|\psi_\lambda\rangle=
E/\lambda^2$, one gets $-2 E/\lambda^3 = \langle \psi_\lambda
|(dH/d\lambda) |\psi_\lambda\rangle=0$, in
contradiction with $E<0$.
\bibitem{abus} There are in general several zero-energy free-space
  eigenstates
with the same exponent $\nu$, so that our notation $\psi_\nu^0$ is
  abusive.
\bibitem{nu_efi}
This holds also when the exponent $\nu$ of the free-space eigenstate
is not real, in which case the Hamiltonian is
not hermitian since the resulting eigenenergy in the trap is complex.
 This happens for the Efimov states of 3 bosons,
which can be written as a sum of scale-invariant states with complex
conjugate exponents.
This examplifies the fact that the present algebra is meaningful for
universal states only.
\bibitem{why_s=0} To obtain Eq.(\ref{eq:s=0}), consider $\psi_2\equiv
  R^\nu u(R) \phi_2(\vec{\Omega})$.
This wavefunction satisfies the contact conditions \cite{lemma}. It
  thus belongs to the domain of $H$ (for an appropriate $u(R)$).
Since $H e_1=0$, hermiticity of $H$ implies $\langle e_1, H \psi_2
\rangle=0$; which leads to the result.
\bibitem{lemma}
To establish this fact, one needs the following lemma:
{\it if $\psi(\vec{X}\,)$ obeys the contact conditions, so does $u(R)
  \psi(\vec{X}\,)$,
where $u(R)$ is a function with no singularity, except maybe in
$R=0$.}
This lemma results from the fact that $R^2= R_0^2 +O(r_{ij}^2)$, where
$R_0^2$ is a non-zero constant in the $r_{ij}~\rightarrow~0$ limit
\cite{lemma2}, so that $u(R)=u(R_0)+O(r_{ij}^2)$.
\bibitem{lemma2}
We assume here $N\geq 3$; in this case,
$R_0$ is different from zero since the contact conditions apply for
$\vec{R}_{ij}\neq\vec{r}_k$,
whatever $k\neq i,j$.
\bibitem{mapping_interne}
The whole algebra of section \ref{sec:mapping} may be reproduced for
the internal problem,
with the modified raising and lowering operators:
${\cal L}_{\pm} \equiv \pm
\left(\frac{3N-3}{2}+R\partial_R\right)+\frac{H_{\rm
    int}}{\hbar\omega} -\frac{m\omega}{\hbar} R^2$.
\bibitem{Nbody_res}
The absence of $N$-body resonance corresponds to the simple boundary
condition: (i) $F(R)$ bounded for
$R\rightarrow 0$.
For $0<s_R<1$, $N$-body resonances can be taken into account by the
modified boundary condition: (ii) $F(R)=A\,R^{s_R}[R^{-2
  s_R}-\epsilon\,
l^{-2 s_R}]+O(R^{-s_R+2})$, where $l\in]0,+\infty]$ and $\epsilon=\pm
  1$ are fixed. If one is not exactly on the $N$-body resonance, one
  has $l<\infty$, which breaks scale invariance, thus invalidating the
  results of this paper. Exactly on the $N$-body resonance
  (e. g. when the energy of the $N$-body bound state in free space
  vanishes),
  one has
  $l=\infty$, which preserves scale invariance: all the results of
  this paper remain valid; one must only replace $s_R$ by $-s_R$ in
  Equations
  (\ref{F_R},\ref{eq:E_int},\ref{F_R_free_space},\ref{psi_int},\ref{nu_s}),
and Eq.(\ref{lower_bound}) becomes $E\geq 3/2 \hbar\omega$.
For $s_R=0$, $N$-body resonances can be taken into account by the
modified boundary condition: (iii)
$F(R)= A \log(R/l) +o(1)$,
where $l\in]0,+\infty[$ is fixed.
If one is not exactly on the $N$-body resonance, one
has $l<\infty$, which breaks scale invariance, thus invalidating the
results of this paper.
Being exactly on the $N$-body resonance corresponds to taking the
limit $l\rightarrow\infty$ in (iii), which
is equivalent to (i): the results of this paper then remain valid.
For $s_R\geq 1$, the wavefunction $F(R)=R^{-s_R}$ is not square
integrable near $R=0$ in $2$ dimensions, so that the description of
$N$-body resonances becomes more complicated (see \cite{Ludo} for the case
$N=2$).
\end{thebibliography}
\end{document}